\documentclass[letterpaper, 12pt]{article}

\usepackage{geometry}
\usepackage{graphicx} 
\usepackage{authblk}
\usepackage{amssymb}
\usepackage{lipsum}
\usepackage{float}
\usepackage{times}
\usepackage[labelfont=bf]{caption}
\usepackage{ragged2e}
\usepackage{longtable}
\usepackage{comment}
\usepackage{csquotes}
\usepackage{setspace}
\usepackage[linkcolor=blue,hyperindex,breaklinks]{hyperref}
\usepackage[russian,ukrainian, english]{babel} 

\usepackage[
backend=biber,
style=numeric-comp,
maxbibnames=250, 
sorting=none,
language=auto,
autolang=other
]{biblatex}

\title{BuyTheBy: A dataset of 18,710 text-based paper mill advertisements with 51,812 timestamped prices}

\author[1*]{Reese AK Richardson}
\author[2]{Spencer S Hong}
\author[3]{Anna Abalkina}
\affil[1]{Center for Science of Science and Innovation, Kellogg School of Management, \break Northwestern University, Evanston, Illinois, USA}
\affil[2]{Department of Engineering Sciences and Applied Mathematics, \break Northwestern University, Evanston, Illinois, USA}
\affil[3]{Institute for East European Studies, Freie Universität Berlin, Berlin, Germany}
\affil[*]{To whom correspondence should be addressed: reese.richardson@northwestern.edu}
\begin{document}

\maketitle

\begin{abstract}
The study of paper mills and similar businesses operating in the market for academic and education fraud services is frustrated by the lack of market price data on their various offerings. Here, we assemble BuyTheBy, a large, annotated dataset of timestamped, text-based paper mill advertisements from seven businesses operating out of seven different countries. The dataset consists of 18,710 individual advertisements, of which 15,839 have prices listed. Among these there are 20,598 positions listed as for sale on 5,567 unique products in 14 different product categories with 51,812 timestamped price data points. We perform elementary analysis of this dataset to demonstrate its utility for quantitative understanding of markets for academic fraud services and suggest future use cases.
\end{abstract}

\section*{Introduction}

Recent years have seen the apparent emergence of “paper mills”, typically understood to be commercial organizations that sell authorship positions on pre-written scientific manuscripts and collude with editors to have these manuscripts published without undergoing a rigorous peer review process \cite{copereport, byrne2022protection, joelving2024paper, brainard2023fake, christopher2021raw}. Paper mill products are suspected to represent an increasing proportion of the published scientific literature \cite{joelving2024paper, van2023big, abalkina2025stamp, richardson2025entities}. We and others have recently suggested that paper mills should be conceptualized as just one aspect of the broader global market for academic fraud services, wherein many actors concurrently offer services related to diploma milling, essay milling, contract cheating, admissions fraud and patent milling \cite{eaton2023fake, ezell2023yesterday, richardson2025exploitation, ogrady2025patent}. Despite ongoing scholarship and journalism enhancing the detection of paper mill products \cite{cabanac2022problematic, richardson2025widespread, porter2024identifying, oste2024misspellings, park2022identification} and describing paper mill operations \cite{joelving2024paper, abalkina2023publication, matusz2025threat, hvistendahl2013china, ro2025authorship, litoy2019scopus, joelving2025broker}, there are many unknowns about paper mill operations and the market for paper mill products \cite{abalkina2025stamp, byrne2024call}. 

One gap in knowledge concerns the market pricing of paper mill products. Thus far, most of the data made available on the market prices of paper mill products has been sparse, anecdotal in nature or indirect. For instance, a 2013 investigation found authorship positions in articles in journals indexed by Web of Science being sold by a Chinese firm for 1,600 United States dollars (USD) to 26,300 USD \cite{hvistendahl2013china}. A Russian company was found to advertise authorship positions for between 180 euros (EUR) to 5,000 EUR between 2019 to 2021 \cite{abalkina2023publication}. A 2024 investigation found that one editor had been offered 3,000 USD for every paper they accepted in their journal originating from one paper mill \cite{joelving2024paper}, implying prices for authorship slots roughly at this level. A 2025 investigation detailing the operations of several paper mills lists prices for authorship positions between 20 USD and 1,500 USD \cite{ro2025authorship}.

To aid in the quantitative economic study of paper mill operations, here we assemble BuyTheBy, dataset of structured price data from largely unstructured text-based paper mill advertisements for a variety of services related to publishing and credential fraud.

\section*{Data and methods}

To assemble BuyTheBy v1.0, we collected advertisements made in ``channels'' on messaging app Telegram from three businesses, using Telegram's ``Export chat history'' feature. We selected these channels because they were publicly accessible channels where many text-based advertisements written mostly in English featuring price data were posted. We prioritized such datasets to assemble as large a dataset as possible with minimal manual cleanup and redaction. Only text-based messages were downloaded; any advertisements contained within PDF or image files attached to messages were not processed. From an additional four businesses, we collected advertisements posted on the paper mill’s website.

\begin{figure}[h!tbp]
    \centering
        \includegraphics[width=\textwidth]{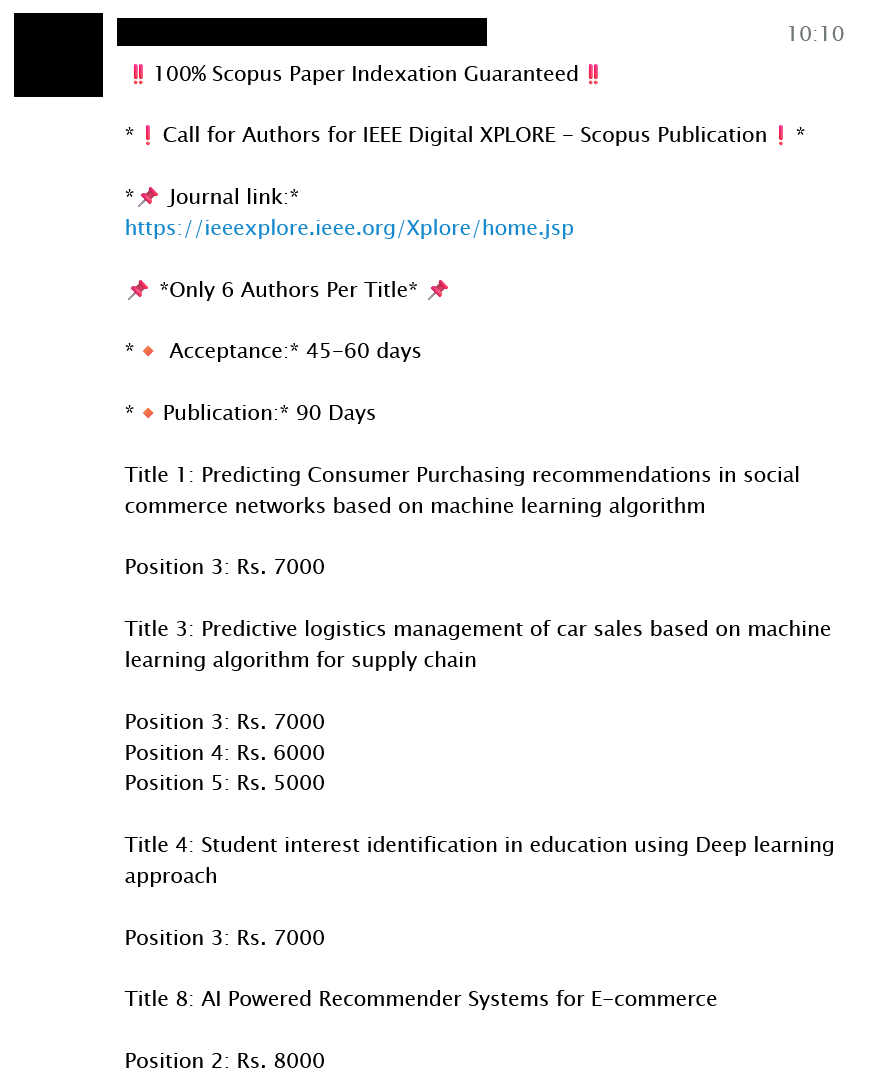}
    \caption{An example advertisement from B1 (posted on 26 June 2024, id\_tag ``message3806'').}
    \label{fig:b1_sample}
\end{figure}

\textbf{Business 1 (B1)} apparently operates in India and primarily caters to academics in India. In its Telegram channel, all messages are written in English and prices are listed in Indian rupees (INR). 1,101 messages were sent in this channel from 16 March 2022 to 16 July 2024. The channel was exported on July 16, 2024, after which the channel owner disabled the “Export chat history feature”. We considered each message to be equivalent to one advertisement. An example advertisement is shown in \textbf{Figure \ref{fig:b1_sample}}.

\begin{figure}[h!tbp]
    \centering
        \includegraphics[width=\textwidth]{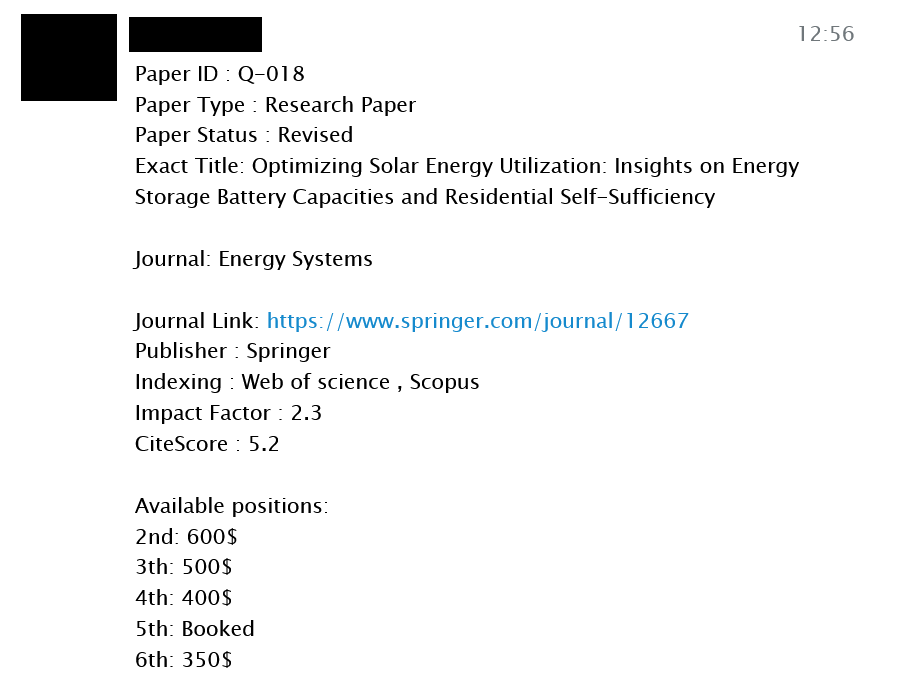}
    \caption{An example advertisement from B2 (posted on 19 March 2024, id\_tag ``message23'').}
    \label{fig:b2_sample}
\end{figure}

\textbf{Business 2 (B2)} apparently operates in Iraq and primarily caters to academics in Iraq. In its Telegram channel, all messages are written in English and prices are listed in USD. Messages spanned two channels, one in which 31 messages were sent by B2 from 20 December 2023 to 26 February 2024 and one in which 53 messages were sent by B2 from 26 February 2024 to 15 June 2024. These channels were exported on 14 August 2024 and 26 June 2024, respectively. We considered each message to be equivalent to one advertisement. An example advertisement is shown in \textbf{Figure \ref{fig:b2_sample}}.

\begin{figure}[h!tbp]
    \centering
        \includegraphics[width=0.75\textwidth]{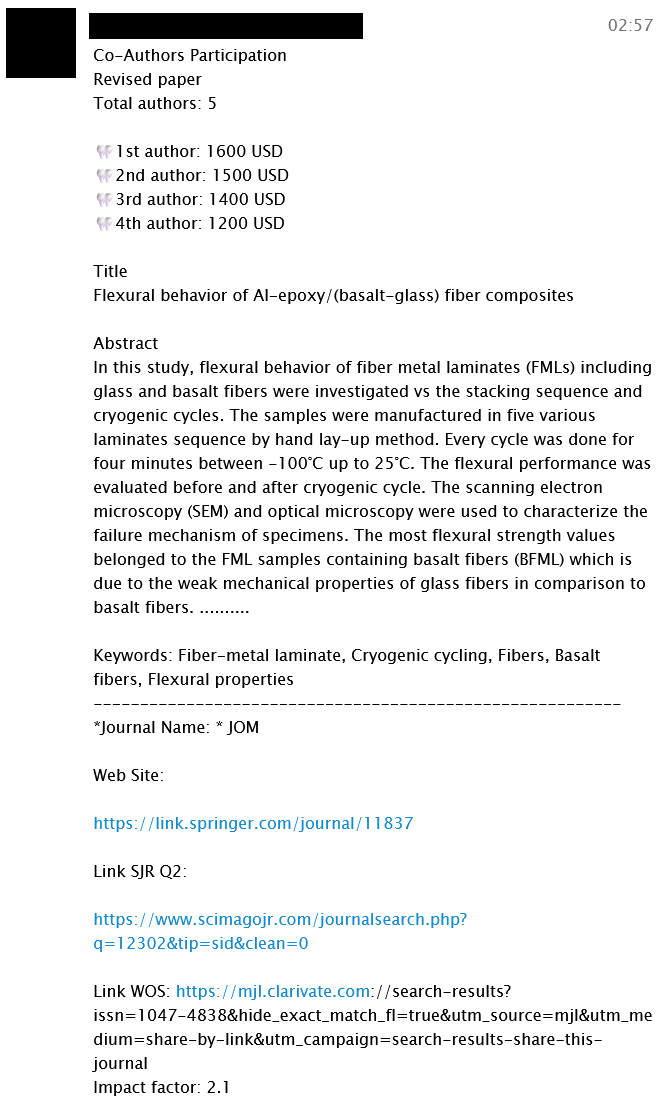}
    \caption{An example advertisement from B3 (posted on 21 March 2025, id\_tag ``message7180'').}
    \label{fig:b3_sample}
\end{figure}

\textbf{Business 3 (B3)} apparently operates in Uzbekistan and primarily caters to academics around Central Asia and the Middle East. In its Telegram channel, most messages are written in English with some messages written in Uzbek. Prices are listed mostly in USD with some prices listed in Uzbekistani sum (UZS). 1,126 messages were sent from 26 March 2020 to 16 March 2026. This channel was exported on 12 March 2025 and 16 March 2026. We considered each message to be equivalent to one advertisement. An example advertisement is shown in \textbf{Figure \ref{fig:b3_sample}}.

\begin{figure}[h!tbp]
    \centering
        \includegraphics[width=\textwidth]{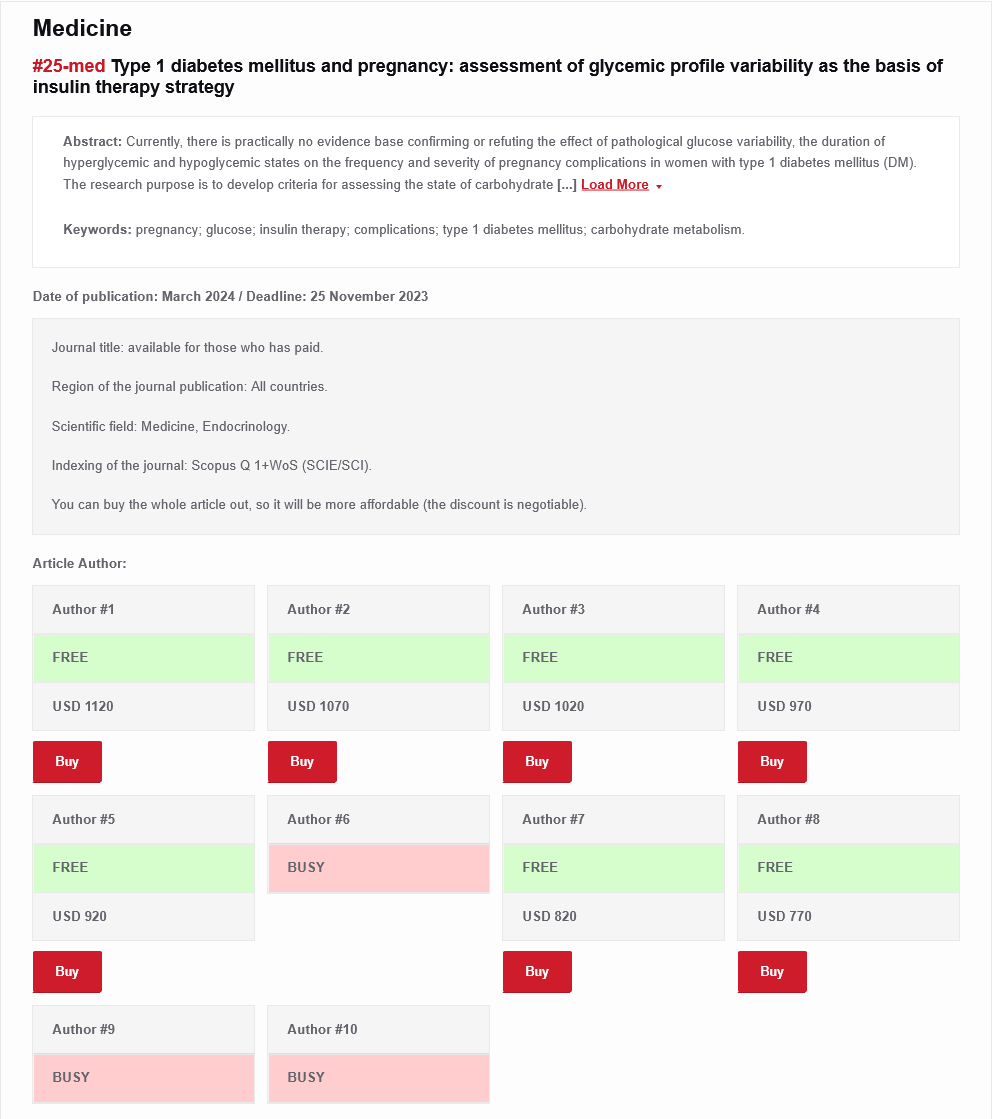}
    \caption{An example advertisement from B4 (archived on 11 November 2023, id\_tag ``231111\_176'').}
    \label{fig:b4_sample}
\end{figure}

\textbf{Business 4 (B4)} is registered in Latvia and appears to cater to academics around Eastern Europe and Central Asia. It has previously been described elsewhere \cite{ro2025authorship}. On its website, B4 lists 178 article topics in English in a searchable table. Each entry lists up to eight authorship slots with prices listed in USD and a button labeled “BUY”, which prompts the user to submit a form with their contact information. We archived the state of this website on 11 November 2023. We considered each entry in this table to be equivalent to one advertisement. An example advertisement is shown in \textbf{Figure \ref{fig:b4_sample}}.

\begin{figure}[h!tbp]
    \centering
        \includegraphics[width=\textwidth]{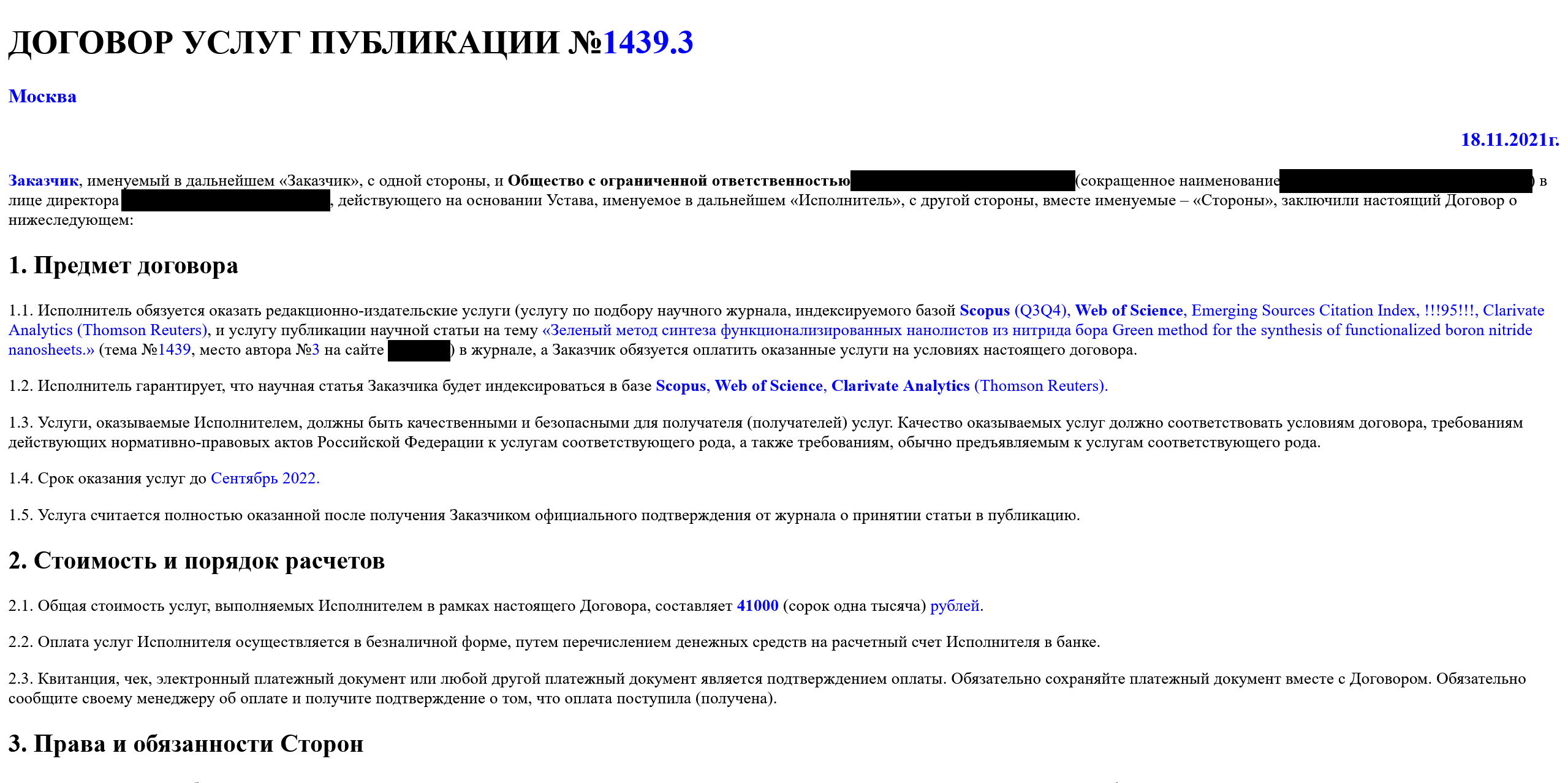}
    \caption{An example advertisement from B5's ``publication service agreement'' pages (archived on 1 November 2021, id\_tag ``1439.3'').}
    \label{fig:b5_sample_0}
\end{figure}

\textbf{Business 5 (B5)} is registered in Russia and appears to primarily cater to academics in Russia and Kazakhstan. It has previously been described elsewhere \cite{abalkina2023publication, litoy2019scopus}. On its former website, B5 hosted “publication service agreement” pages that list an article topic, an authorship position on that article and a price for this single authorship slot in Russian rubles (RUB). A button on the bottom of each contract redirects the user to a payment page. We archived the state of these pages on 1 October 2021. However, prior to this date, B5 had become aware that its services were being monitored by us and added 17,763 fake contracts where article titles were derived from previously published articles. We were able to discriminate between fake and real contracts because fake contracts did not have a functional purchase button and uniformly listed the price of advertised authorship slot as zero. We exclude these contracts from our dataset. We considered each of the remaining 7,907 pages, with one authorship slot each, to be equivalent to one advertisement. An example advertisement is shown in \textbf{Figure \ref{fig:b5_sample_0}}.

\begin{figure}[h!tbp]
    \centering
        \includegraphics[width=\textwidth]{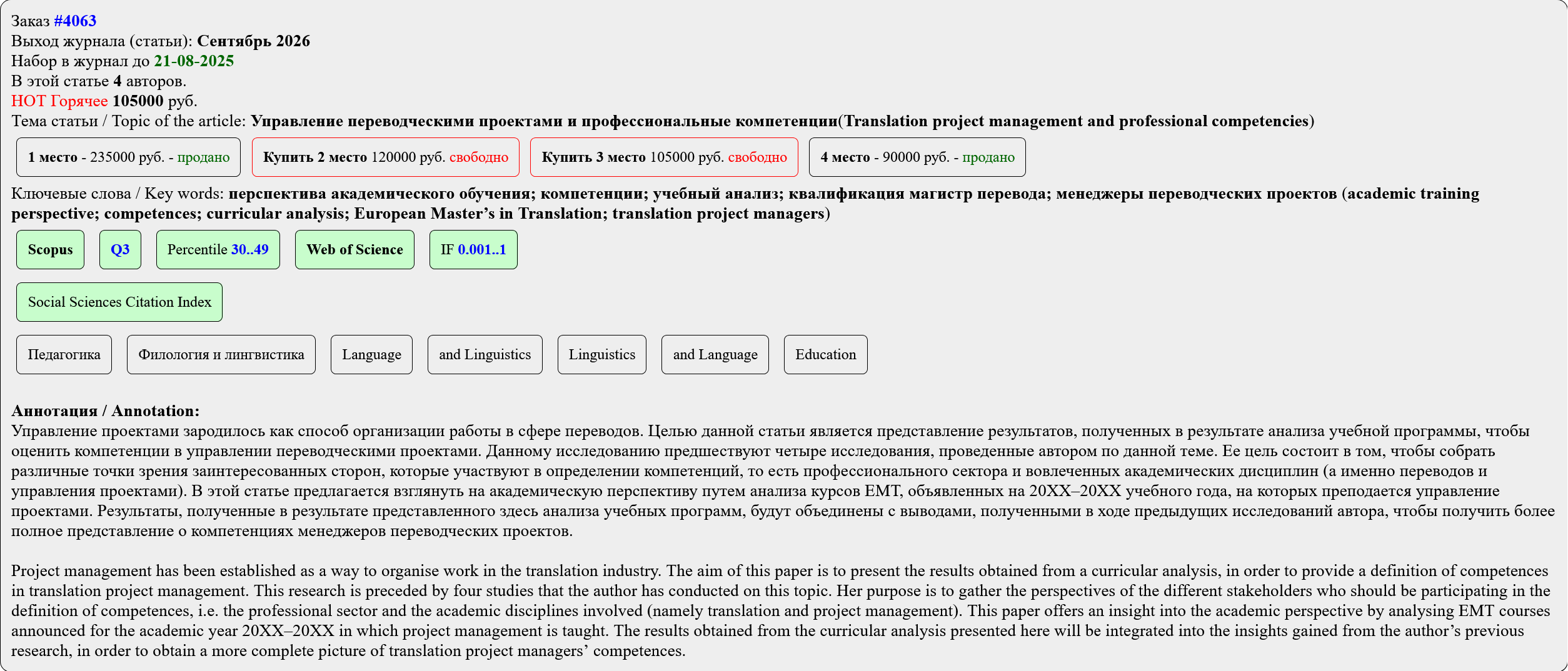}
    \caption{An example advertisement from B5's new wewsbite pages (archived on 7 August 2025, id\_tag ``4063\_250807'').}
    \label{fig:b5_sample_1}
\end{figure}

On a different, newer website, B5 lists article topics and characteristics of target journals alongside available authorship slots with prices in RUB. We archived the state of this page on 13 dates between 14 April 2024, when it listed 85 topics, and 4 April 2026, when it listed 491 topics. We considered each listed topic to be one advertisement. An example advertisement is shown in \textbf{Figure \ref{fig:b5_sample_1}}.

\begin{figure}[h!tbp]
    \centering
        \includegraphics[width=0.75\textwidth]{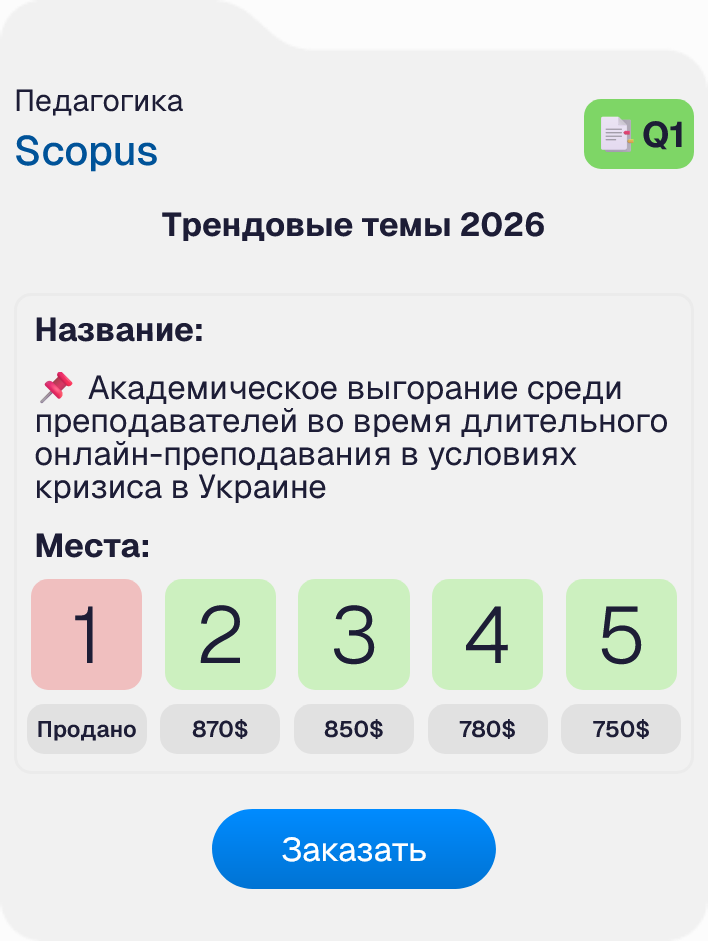}
    \caption{An example advertisement from B6 (archived on 21 March 2026, id\_tag ``260321 \selectlanguage {russian} Педагогика/Образование \selectlanguage{english} 43'').}
    \label{fig:b6_sample}
\end{figure}

\textbf{Business 6 (B6)} apparently operates in Ukraine and primarily caters to academics in Ukraine. B6's website, available in English, Ukrainian and Russian, has pages advertising authorship slots on articles, textbooks and ``abstracts'' (which we take to refer abstracts in conference proceedings). Prices are listed in USD and Ukrainian hryvnias (UAH). The pages advertising articles were organized by field (e.g., ``Pedagogy/Education'', ``Medical Sciences'', etc.). We archived the state of the Russian version of these pages on 7 September 2024, 13 December 2025 and 21 March 2026 (on 7 September 2024, the ``Humanities'' page was only available in Ukrainian). We archived the state of the Russian version of the page advertising textbooks on six dates between 7 October 2024 and 25 March 2026. We archived the state of the Ukrainian version of the page advertising abstracts on five dates between 23 August 2024 and 1 April 2026. We decided to process the Russian version of the article pages, the Ukrainian version of the textbook page and the Russian version of the abstract pages because these were the languages in which the greatest number of past versions of each page were available on webpage archiving services. Many topics are listed in English even on the Russian and Ukrainian versions of each product page. We consider each listed topic to be one advertisement. An example advertisement is shown in \textbf{Figure \ref{fig:b6_sample}}.

\begin{figure}[h!tbp]
    \centering
        \includegraphics[width=0.75\textwidth]{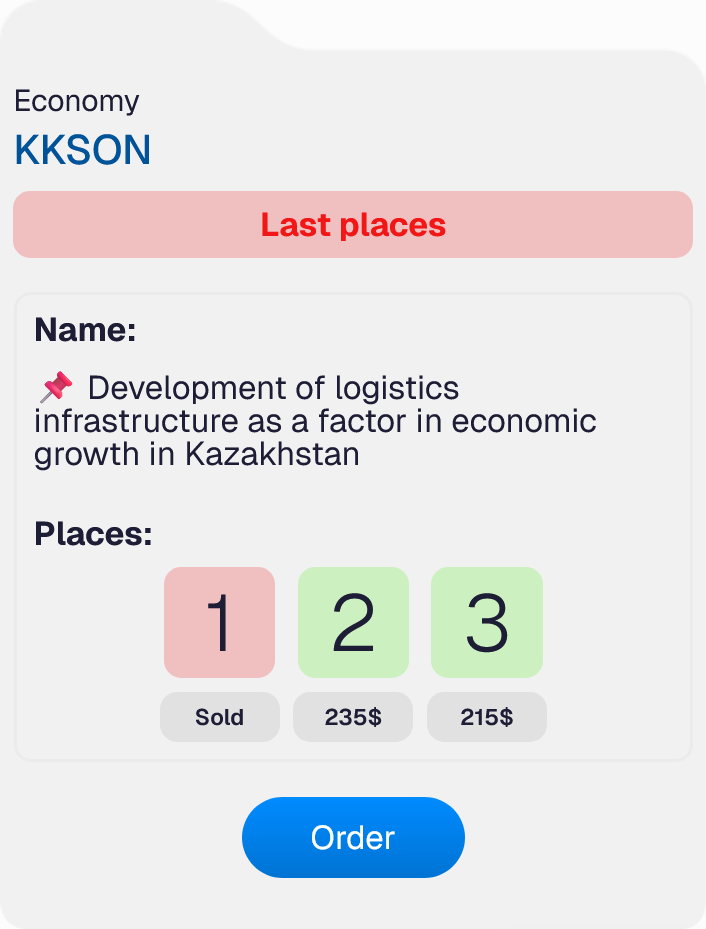}
    \caption{An example advertisement from B7 (archived on 21 March 2026, id\_tag `260321 Economic Sciences 5'').}
    \label{fig:b7_sample}
\end{figure}

\textbf{Business 7 (B7)} apparently operates in Kazakhstan and primarily caters to academics in Kazakhstan. B7's website, available in English, Kazakh and Russian, features nearly identical branding and layout to B6's website. Despite this, the two sites list different addresses (in Ukraine and Kazakhstan, respectively) and phone numbers, link to different social media profiles and do not link to one another. We consider B6 and B7 to be separate paper mills because their offerings are largely distinct. B7's website only advertises authorship slots on articles, also organized by field. We archived the state of the English version of these pages on 28 March 2025 and 21 March 2026. Prices are listed in USD. We consider each listed topic to be one advertisement. An example advertisement is shown in \textbf{Figure \ref{fig:b7_sample}}.

Metadata and prices from these advertisements were obtained using a combination of automated and manual extraction and validation. Product titles were only extracted from advertisements featuring prices. To allow for qualitative analysis of these datasets, we provide the text of these advertisements with business names, websites, phone numbers, social media links and individuals’ names redacted. These redactions were made with custom regular expressions. Prices were converted to USD based on historical exchange rate data provided by the United States Department of the Treasury \cite{exchangerates}.

The processed BuyTheBy dataset is available on \href{https://doi.org/10.5281/zenodo.19684277}{Zenodo} \cite{buythebyzenodo} under a CC BY 4.0 license alongside a README file and data dictionary describing data fields. Code for reproducing figures and summary statistics is available at \href{https://github.com/reeserich/buytheby}{github.com/reeserich/buytheby} \cite{buythebygithub}.

\section*{Summary and analysis}

Summary data for advertisements from each business are provided in \textbf{Table \ref{tab:summary_all}}.

\begin{table}[h!tbp]
    \begin{center}
        \caption{Summary data for collected paper mill advertisements. Two products were considered the same unique product if they were advertised by the same business in the same product category and their titles matched exactly (insensitive to case or punctuation).}
        \label{tab:summary_all}
        \begin{tabular}{ |p{2cm}|p{1.6cm}|p{1.6cm}|p{1.6cm}|p{1.6cm}|p{1.6cm}|p{1.6cm}|p{1.6cm}| } 
         \hline
         \textbf{Business} & \textbf{B1} & \textbf{B2} & \textbf{B3} & \textbf{B4} & \textbf{B5} & \textbf{B6} & \textbf{B7} \\ \hline \hline
         \textbf{Platform} & Telegram & Telegram & Telegram & Website & Website & Website & Website \\ \hline
         \textbf{Earliest advertisement} & 16 Mar 2022 & 20 Dec 2023 & 26 Mar 2020 & 11 Nov 2023 & 1 Oct 2021 & 23 Aug 2024 & 28 Mar 2025\\ \hline
         \textbf{Latest advertisement} & 16 Jul 2024 & 15 Jun 2024 & 16 Mar 2026 & 11 Nov 2023 & 4 Apr 2026 & 1 Apr 2026 & 21 Mar 2026\\ \hline
         \textbf{Number of advertisements} & 1,101 & 84 & 1,126 & 178 & 12,915 & 2,957 & 349\\ \hline
         \textbf{Product types listed} & 13 & 1 & 1 & 1 & 1 & 3 & 1\\ \hline
         \textbf{Number of advertisements with price(s) listed} & 694 & 49 & 332 & 178 & 11,283 & 2,954 & 349\\ \hline
         \textbf{Number of unique products listed with price} & 692 & 26 & 259 & 178 & 2,543 & 1,605 & 264\\ \hline
         \textbf{Number of unique positions listed with price} & 2,061 & 102 & 779 & 784 & 9,551 & 6,109 & 1,212\\ \hline
         \textbf{Number of timestamped price data points} & 15,578 & 142 & 952 & 785 & 21,089 & 11,678 & 1,588\\ \hline
        \end{tabular}
    \end{center}
\end{table}

Each business’ advertisements differ markedly from one another. For instance, while B1 and B6 advertise a diversity of products, including authorship positions on academic articles and textbooks, editor positions on textbooks and inventorship positions on patents, the remaining five businesses only advertise authorship positions on academic articles as for sale. B1's offerings are summarized in \textbf{Table \ref{tab:b1}}

\begin{table}[h!tbp]
    \begin{center}
        \caption{Types of products offered by B1. Only advertisements/products/positions with prices are listed. Products are considered the same unique product if they are in the same product category and  their titles matched exactly (insensitive to case or punctuation).}
        \label{tab:b1}
        \begin{tabular}{ |p{4cm}|p{2.5cm}|p{2.5cm}|p{2.5cm}|p{2.5cm}| } 
         \hline
         \textbf{Type of product} & \textbf{Number of advertisements} & \textbf{Number of unique products} & \textbf{Number of unique positions} & \textbf{Number of timestamped prices}\\ \hline \hline
         \textbf{Authorship of ``international'' textbook} & 234 & 257 & 637 & 6,865 \\ \hline
         \textbf{Authorship of article} & 99 & 88 & 210 & 1,667 \\ \hline
         \textbf{India utility patent} & 98 & 80 & 522 & 3,143 \\ \hline
         \textbf{Editorship of ``international'' textbook} & 88 & 28 & 141 & 604 \\ \hline
         \textbf{United Kingdom design registration (advertised as ``UK design patent'' \cite{richardson2025exploitation})} & 41 & 18 & 112 & 615 \\ \hline
         \textbf{India copyright registration} & 29 & 6 & 19 & 108 \\ \hline
         \textbf{India design registration (advertised as ``India design patent'' \cite{richardson2025exploitation})} & 28 & 15 & 63 & 210 \\ \hline
         \textbf{Canada copyright registration} & 21 & 9 & 54 & 492 \\ \hline
         \textbf{Authorship of ``international'' textbook chapter} & 21 & 173 & 173 & 955 \\ \hline
         \textbf{Germany utility patent} & 18 & 3 & 24 & 79 \\ \hline
         \textbf{South Africa utility patent} & 8 & 1 & 8 & 55 \\ \hline
         \textbf{Australia design registration (advertised as ``Australia design patent'' \cite{richardson2025exploitation})} & 7 & 12 & 96 & 680 \\ \hline
         \textbf{``National'' award} & 2 & 2 & 2 & 4 \\ \hline
        \end{tabular}
    \end{center}
\end{table}

Among advertisements for academic articles, B1, B2 and B3 typically name the target journal, while B4, B5, B6 and B7 never name the target journal but do frequently indicate its indexing status and approximate rank (e.g., ``Scopus Q1'', ``WoS Emerging Sources Citation Index Q4”, ``Category B''). While B2 and B3 advertise authorship positions on articles that they imply will be published in indexed journals published by large, well-regarded international publishers like Springer Nature and Elsevier, B1 almost exclusively advertises authorship positions in Institute of Electrical and Electronics Engineers (IEEE) conference proceedings and articles in journals belonging to small, regional publishers, many of which have likely been hijacked \cite{abalkina2023challenges, abalkina2021detecting}. The representation of publishers among advertisements that do name the target venue is summarized in \textbf{Table \ref{tab:publishers}}. The representation of selected indexing services and journal lists among all advertisements is summarized in \textbf{Table \ref{tab:indexing}}.

\begin{table}[h!tbp]
    \begin{center}
        \caption{Representation of publishers among advertisements that name the targeted publication venue.}
        \label{tab:publishers}
        \begin{tabular}{ |p{6cm}|p{2cm}|p{2cm}| } 
         \hline
         \textbf{Publisher} & \textbf{Number of advertisements} & \textbf{Number of unique products listed with price} \\ \hline \hline
         \textbf{Elsevier} & 112 & 76 \\ \hline
         \textbf{IEEE} & 101 & 54 \\ \hline
         \textbf{Springer Nature} & 86 & 78 \\ \hline
         \textbf{European Economics Letters} & 22 & 6 \\ \hline
         \textbf{Wiley} & 22 & 20 \\ \hline
         \textbf{Lutheran University of Brazil} & 20 & 7 \\ \hline
         \textbf{Taylor \& Francis} & 14 & 9 \\ \hline
         \textbf{MDPI} & 8 & 4 \\ \hline
         \textbf{De Gruyter Brill} & 8 & 5 \\ \hline
         \textbf{Royal Society of Chemistry} & 7 & 6 \\ \hline
         \textbf{Other} & 74 & 100 \\ \hline
        \end{tabular}
    \end{center}
\end{table}

\begin{table}[h!tbp]
    \begin{center}
        \caption{Representation of select indexing services and journal lists among all advertisements. Advertisements either contained the name of the service (e.g., ``Web of Science''), a sub-index (e.g., ``Arts \& Humanities Citation Index'') or a metric closely related to the service (e.g., ``Impact factor'').}
        \label{tab:indexing}
        \begin{tabular}{ |p{7cm}|p{3.5cm}|p{3.5cm}| } 
         \hline
         \textbf{Indexing service or journal list} & \textbf{Number of advertisements} & \textbf{Number of unique products listed with price} \\ \hline \hline
         \textbf{Scopus} & 12,012 & 3,220 \\ \hline
         \textbf{Web of Science} & 5,136 & 1,294 \\ \hline
         \textbf{Category B (Ukraine)} & 97 & 96 \\ \hline
         \textbf{Australian Business Deans Council (ABDC) Journal Quality List} & 27 & 6 \\ \hline
         \textbf{PubMed/MEDLINE} & 21 & 10 \\ \hline
         \textbf{KKSON (Kazakhstan)} & 15 & 11 \\ \hline
         \textbf{UGC CARE list (India, discontinued)} & 12 & 0 \\ \hline
        \end{tabular}
    \end{center}
\end{table}

From here, we provide some elementary exploratory data analysis to demonstrate the BuyTheBy’s potential utility.

\begin{figure}[h!tbp]
    \centering
        \includegraphics[width=\textwidth]{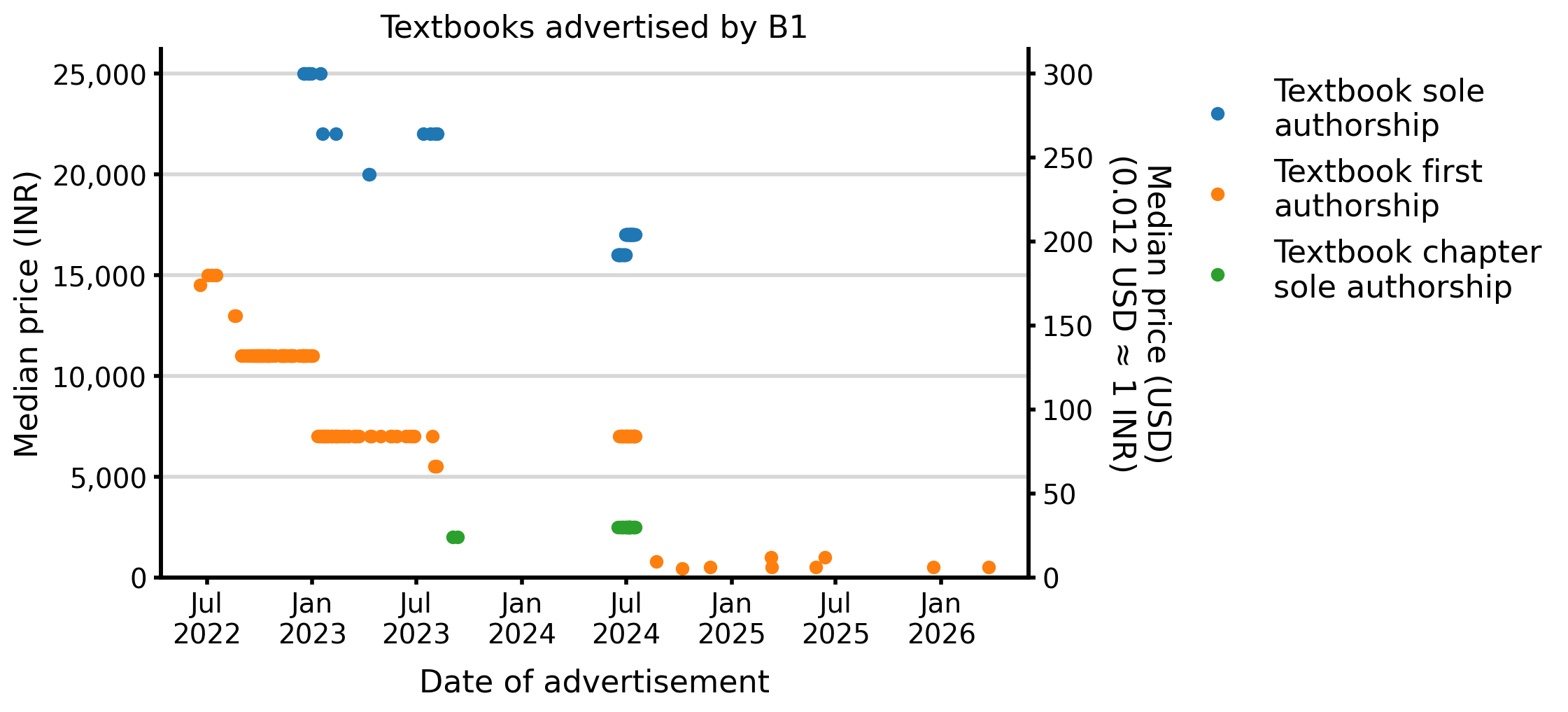}
    \caption{The median listed price for sole authorship, first authorship and sole authorship of a single chapter of an ``international'' textbook changed over time in advertisements posted by B1. Prices were converted from INR to USD at a static exchange rate of 0.012 USD to 1 INR.}
    \label{fig:b1_textbook}
\end{figure}

\begin{figure}[h!tbp]
    \centering
        \includegraphics[width=\textwidth]{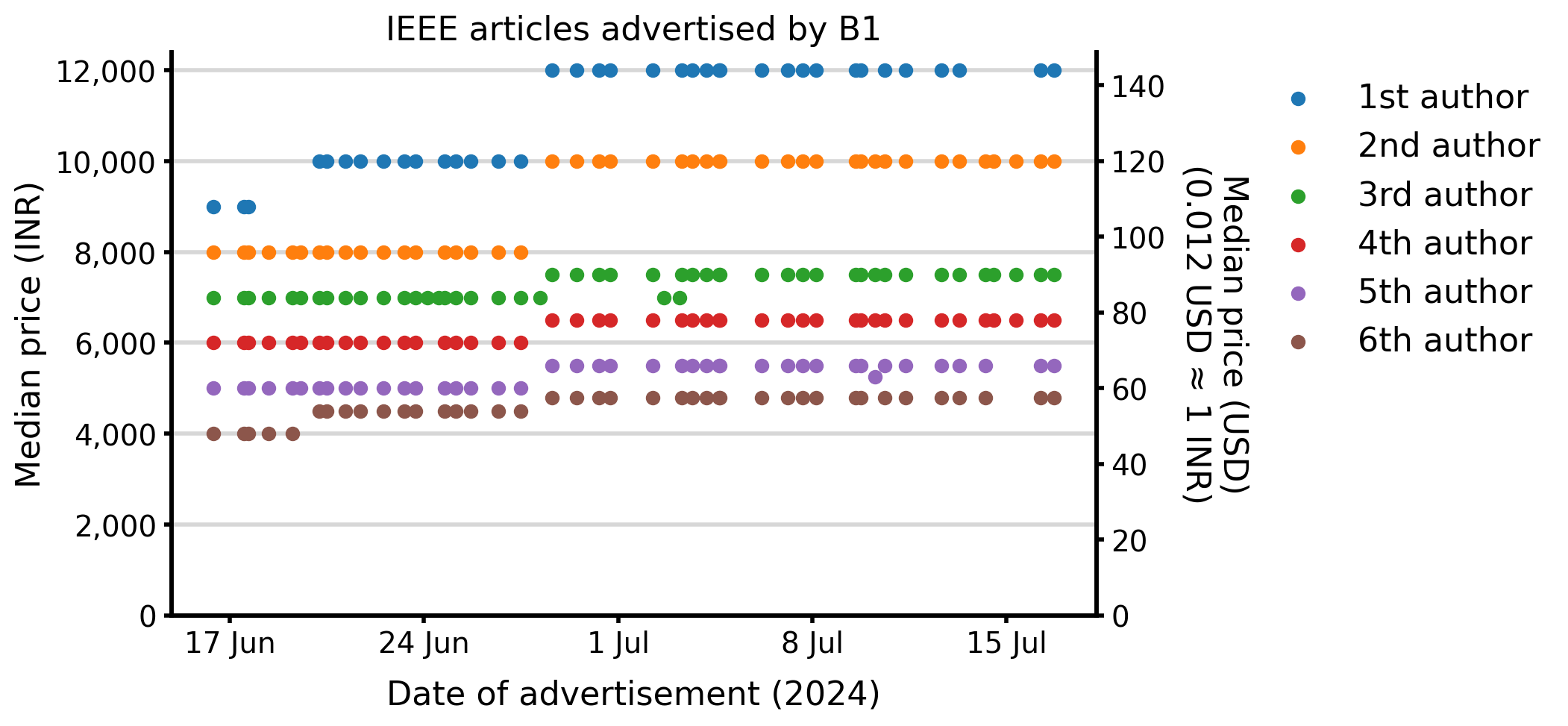}
    \caption{The median listed price for authorship positions on IEEE conference proceeding articles changed over time in advertisements posted by B1. Prices were converted from INR to USD at a static exchange rate of 0.012 USD to 1 INR.}
    \label{fig:b1_ieee}
\end{figure}

One unique feature of this dataset is that it allows for comparison of movement in market prices over time. For instance, between June and December 2022, the median price advertised for first authorship on an ``international'' textbook by B1 dropped several times from about 15,000 INR (about 180 USD) to about 7,000 INR (about 84 USD), as shown in \textbf{Figure \ref{fig:b1_textbook}}. Similarly, in the single month when B1 was advertising authorship positions on IEEE conference proceeding articles captured in our dataset (in June/July 2024), the going rate for each authorship position apparently increased twice, as shown in \textbf{Figure \ref{fig:b1_ieee}}.

\pagebreak

\begin{figure}[h!tbp]
    \centering
        \includegraphics[width=\textwidth]{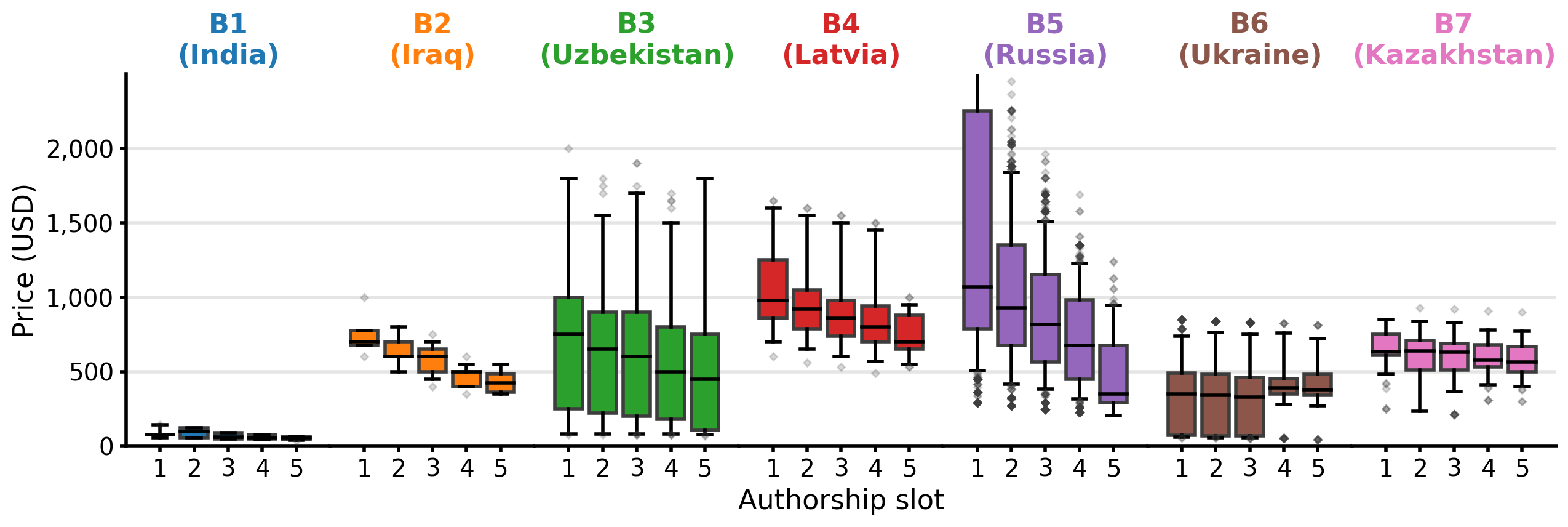}
    \caption{Distributions of prices for authorship positions on academic articles advertised by each business. Only the most recently-advertised price for a given authorship position is included and only authorship slots one through five are shown. Boxplots show median as a horizontal line, interquartile range as boxes, 2.5th and 97.5th percentiles as whiskers, and outliers as diamonds. Prices were converted to USD based on historical exchange rate data provided by the United States Department of the Treasury \cite{exchangerates}. The y-axis is truncated at 2,500 USD. B5 advertised 438 first author positions and three other positions above this price. The most expensive price listed by B5 was 5,631 USD. Extended versions of this figure are shown in \textbf{Figure \ref{fig:b1_all}} through \textbf{\ref{fig:b7_all}}.}
    \label{fig:all_prices}
\end{figure}

This dataset also allows for comparison of prices for similar products in different markets, as demonstrated for authorship positions on academic articles among the seven businesses included in the dataset in \textbf{Figure \ref{fig:all_prices}}. Although our selection of markets is limited, it is clear that prices vary considerably between businesses and countries, likely as a function of the income level of each business' targeted clientele.

Context-dependence, extreme variability in prices and unclear representativeness notwithstanding, this dataset does allow for central estimates for prices to be made for the first time. Across all advertisements included in this dataset, the median price for a first author position on an academic article is 788 USD (see other central estimates in \textbf{Table \ref{tab:estimates}}).

\begin{table}[h!tbp]
    \begin{center}
        \caption{Descriptive statistics of prices per authorship slot on academic articles (in USD). Estimates are taken across all advertisements from all seven businesses. Only the most recently-advertised price for a given authorship position is included.}
        \label{tab:estimates}
        \begin{tabular}{ |p{3cm}|p{1.5cm}|p{1.5cm}|p{1.5cm}|p{1.5cm}|p{1.5cm}| } 
         \hline
        \textbf{Authorship slot}  & \textbf{Count} & \textbf{Median} & \textbf{Mean} & \textbf{Min}  & \textbf{Max} \\ \hline \hline
        \textbf{1}       & 4,206 & 788.41 & 1,031.85 & 56.87 & 5,631.48 \\ \hline
        \textbf{2}       & 4,376 & 675.78  & 760.88  & 53.985 & 5,068.33                            \\ \hline
        \textbf{3}       & 4,397 & 570.00      & 676.81 & 47.987 & 4,505.19 \\ \hline
        \textbf{4}       & 2,995 & 560.00      & 636.48 & 41.99   & 3,378.89 \\ \hline
        \textbf{5}       & 1,618 & 420.00      & 464.47 & 36.00   & 1,800.00  \\ \hline
        \textbf{6}       & 154 & 462.76      & 460.67 & 48.00   & 1,550.00   \\  \hline
        \textbf{7}       & 35 & 450.00      & 514.71 & 200.00      & 1,500.00 \\  \hline
        \textbf{8}       & 28 & 450.00      & 539.11 & 250.00      & 1,650.00     \\   \hline
        \textbf{9}       & 16 & 350.00      & 454.69 & 150.00      & 1,200.00    \\  \hline
        \textbf{10}      & 5 & 250.00      & 246.00     & 130.00      & 350.00 \\ \hline
        \end{tabular}
    \end{center}
\end{table}

\section*{Discussion}

Here, we provide a first-of-its-kind annotated dataset of paper mill advertisements. Elementary analysis of the dataset confirms that having such rich datasets is important for the quantitative understanding of paper mill operations: prices can fluctuate considerably over time and vary wildly across different markets and contexts, even within a single business' portfolio of offerings.

Future studies using BuyTheBy might interrogate how prices vary based on the perceived quality of product being offered (e.g., indexing status or ranking of a journal), based on the perceived reward for purchasing the product (e.g., by linking the dataset with corresponding points-based metrics in widely adopted institutional tenure and promotion criteria) or based on academic discipline. Publishers and indexing services might also use BuyTheBy to augment internal assessments of their portfolios’ exposure to paper mill operations (similar to how previous works have linked paper mill advertisements to eventually published articles \cite{abalkina2023publication}). BuyTheBy also provides opportunities for in-depth qualitative study of the advertising and communications strategies employed by paper mills.

This dataset does bear some limitations. First, we only extract text-based advertisements and make no attempt to extract information from any image- or video-based advertisements posted in these or other channels or websites. Second, our selection of which channels or websites to include is not systematic, having mostly been motivated by informal selection criteria intended to ease assembly and distribution of the dataset. Third, our dataset has limited coverage, geographically and in time. Finally, we make no attempt to link any of these advertisements to products that were eventually published (although many advertisements contain sufficient information to do so). Those interested in making these links should be aware that paper mills are known to change the text of articles from the posted advertisement to frustrate efforts to identify published paper mill products and protect clients \cite{ro2025authorship}. Moreover, paper mill products that are eventually published often do not appear in the same venue that was initially advertised. Finally, users should note that advertised topics are often modeled after or directly plagiarize previously-published articles.

\section*{Acknowledgements}

We thank Nick Wise for providing data and feedback for this project and participants in the 2025 Calgary Colloquium for valuable discussions. The play on words ``Buy the `By''' was borrowed from the title of \textit{Nature}'s print version of a 2025 investigation by Christine Ro and Jack Leeming \cite{ro2025authorship}. RR contributed to this work independently of his work as a postdoctoral researcher at Northwestern University.

\section*{Data availability}

The processed BuyTheBy dataset is available on \href{https://doi.org/10.5281/zenodo.19684277}{Zenodo} \cite{buythebyzenodo} alongside a README file and data dictionary describing data fields. 

\section*{Code availability} 

Code for reproducing figures and summary statistics is available at \href{https://github.com/reeserich/buytheby}{github.com/reeserich/buytheby} \cite{buythebygithub}.

\section*{Abbreviations}

United States dollars (USD); Euro (EUR); Indian rupees (INR); Ukrainian hryvnias (UAH); Uzbekistani sum (UZS); Russian rubles (RUB); Institute of Electrical and Electronics Engineers (IEEE).

\printbibliography


\begin{figure}[h!tbp]
    \centering
        \includegraphics[width=0.75\textwidth]{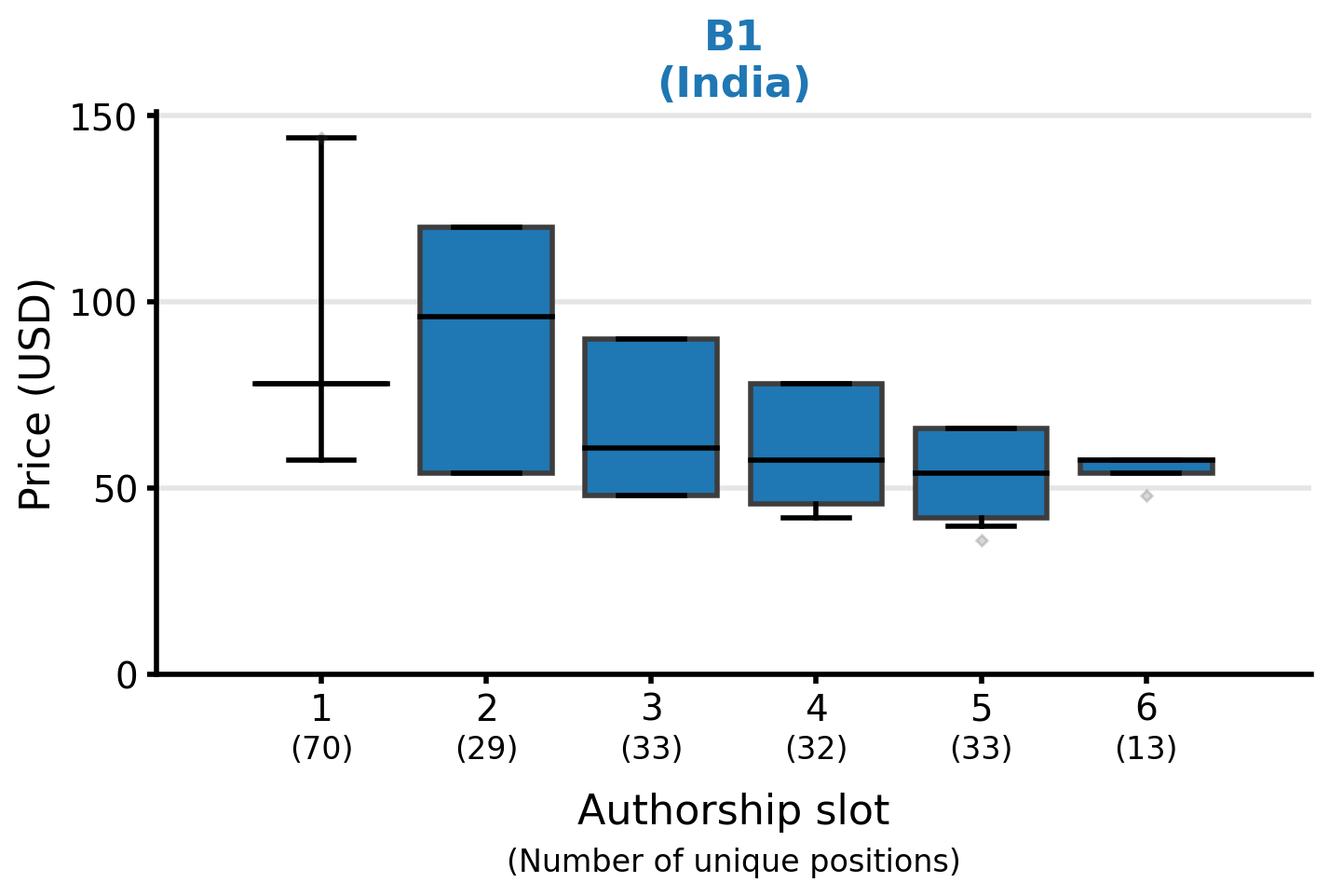}
    \caption{Distributions of prices for authorship positions on academic articles advertised by B1. Only the most recently-advertised price for a given authorship position is included. Boxplots show median as a horizontal line, interquartile range as boxes, 2.5th and 97.5th percentiles as whiskers, and outliers as diamonds.}
    \label{fig:b1_all}
\end{figure}

\begin{figure}[h!tbp]
    \centering
        \includegraphics[width=0.75\textwidth]{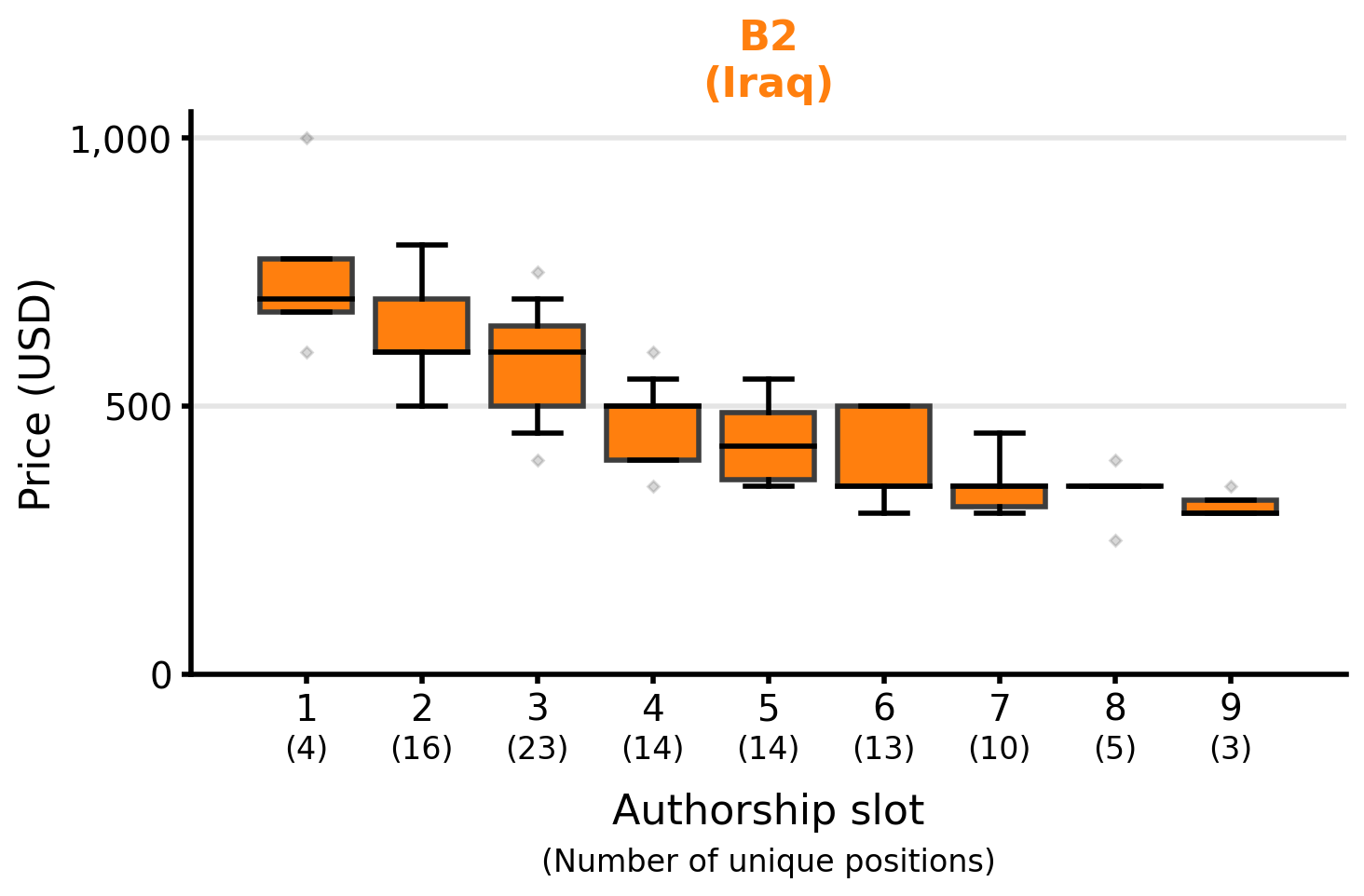}
    \caption{Distributions of prices for authorship positions on academic articles advertised by B2. Only the most recently-advertised price for a given authorship position is included. Boxplots show median as a horizontal line, interquartile range as boxes, 2.5th and 97.5th percentiles as whiskers, and outliers as diamonds.}
    \label{fig:b2_all}
\end{figure}

\begin{figure}[h!tbp]
    \centering
        \includegraphics[width=0.75\textwidth]{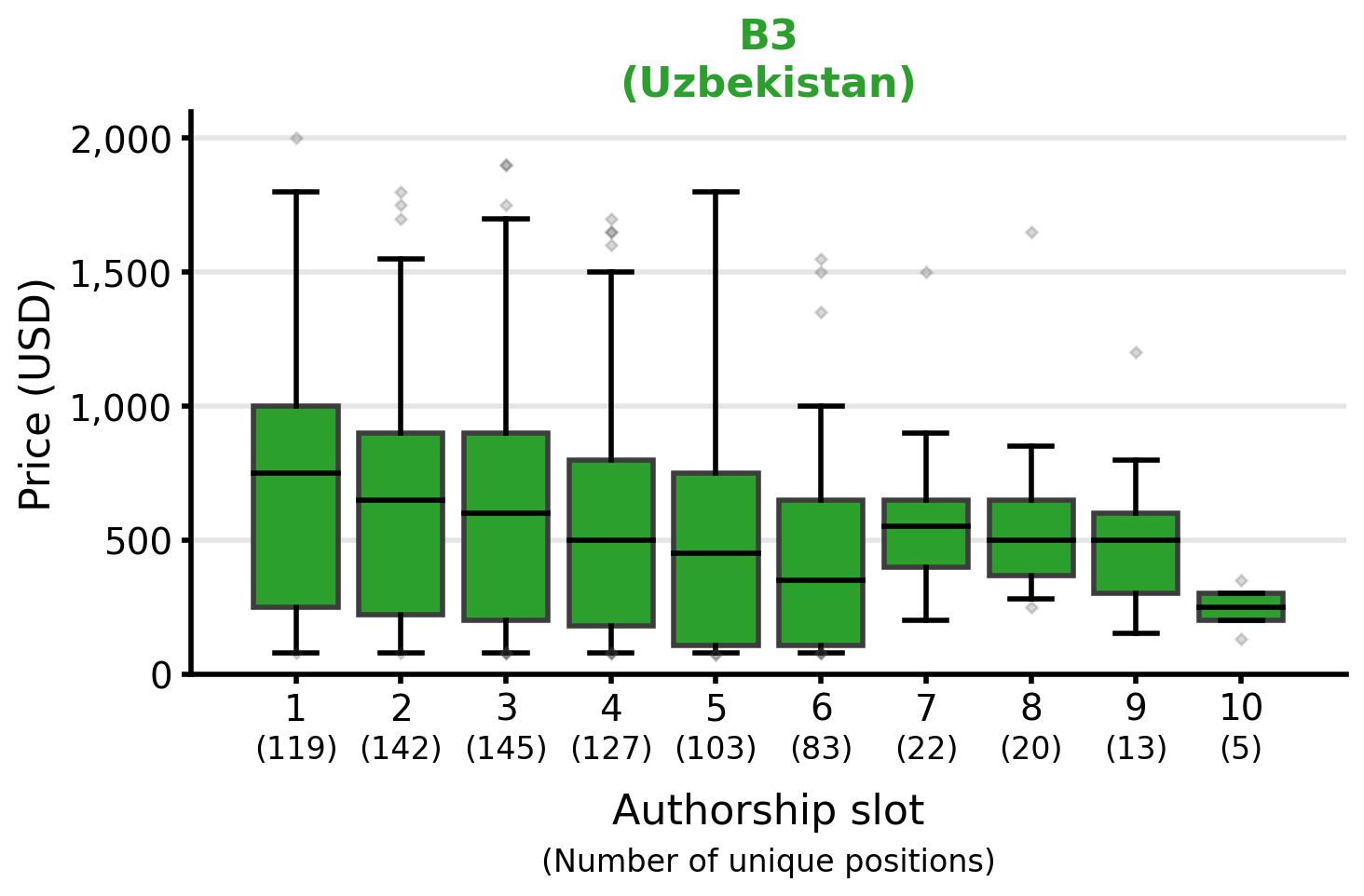}
    \caption{Distributions of prices for authorship positions on academic articles advertised by B3. Only the most recently-advertised price for a given authorship position is included. Boxplots show median as a horizontal line, interquartile range as boxes, 2.5th and 97.5th percentiles as whiskers, and outliers as diamonds.}
    \label{fig:b3_all}
\end{figure}

\begin{figure}[h!tbp]
    \centering
        \includegraphics[width=0.75\textwidth]{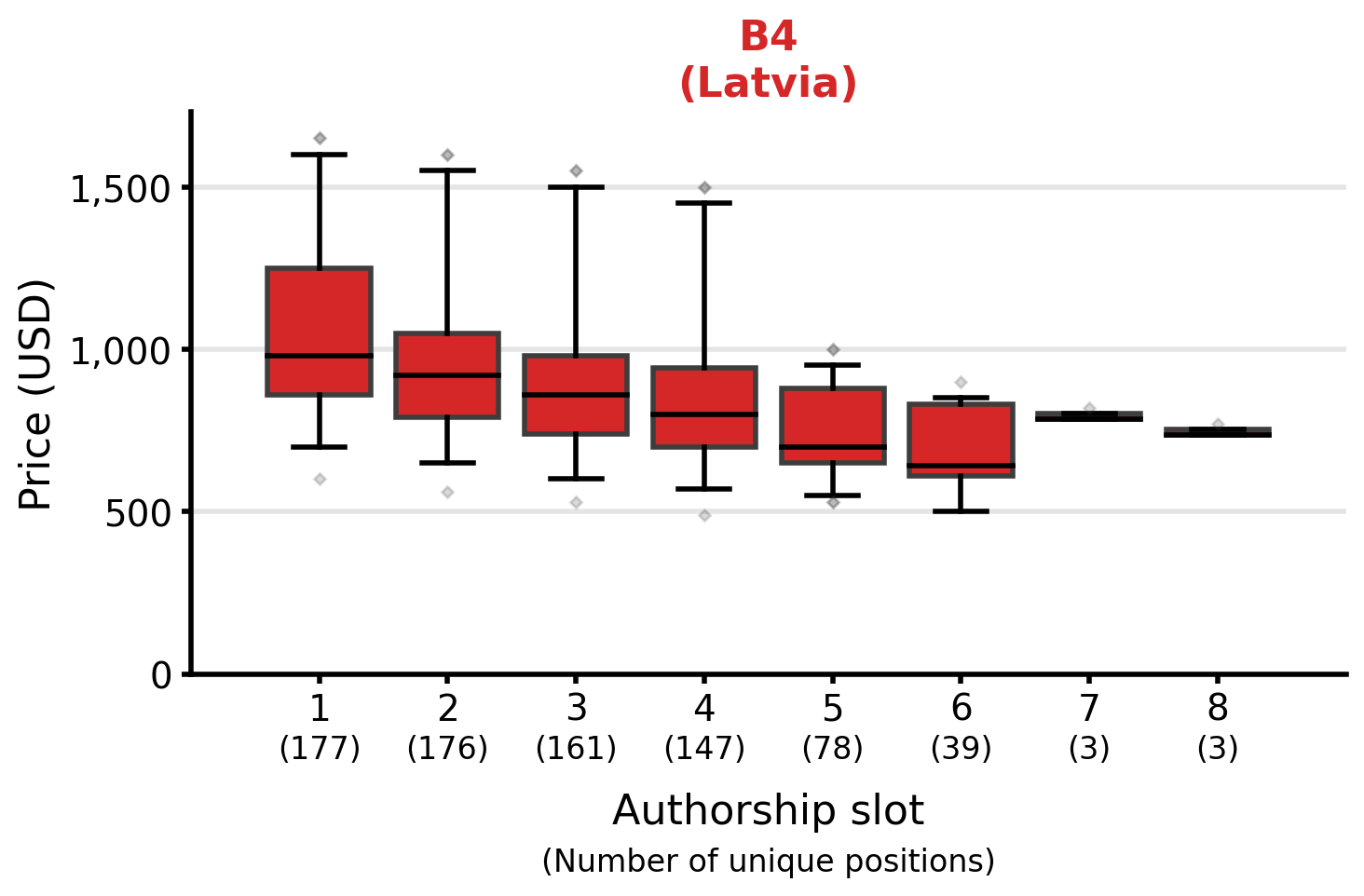}
    \caption{Distributions of prices for authorship positions on academic articles advertised by B4. Only the most recently-advertised price for a given authorship position is included. Boxplots show median as a horizontal line, interquartile range as boxes, 2.5th and 97.5th percentiles as whiskers, and outliers as diamonds.}
    \label{fig:b4_all}
\end{figure}

\begin{figure}[h!tbp]
    \centering
        \includegraphics[width=0.75\textwidth]{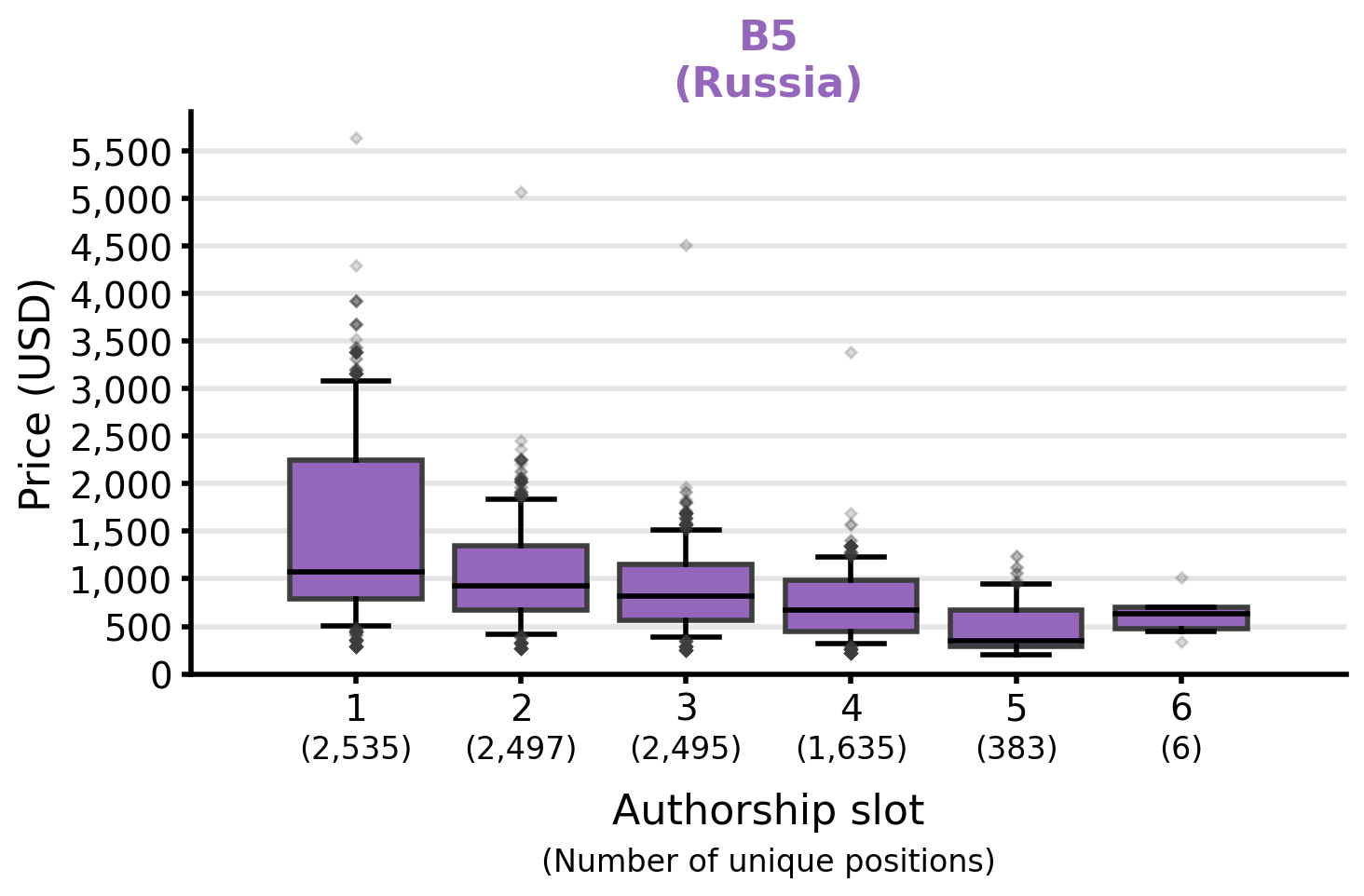}
    \caption{Distributions of prices for authorship positions on academic articles advertised by B5. Only the most recently-advertised price for a given authorship position is included. Boxplots show median as a horizontal line, interquartile range as boxes, 2.5th and 97.5th percentiles as whiskers, and outliers as diamonds.}
    \label{fig:b5_all}
\end{figure}

\begin{figure}[h!tbp]
    \centering
        \includegraphics[width=0.75\textwidth]{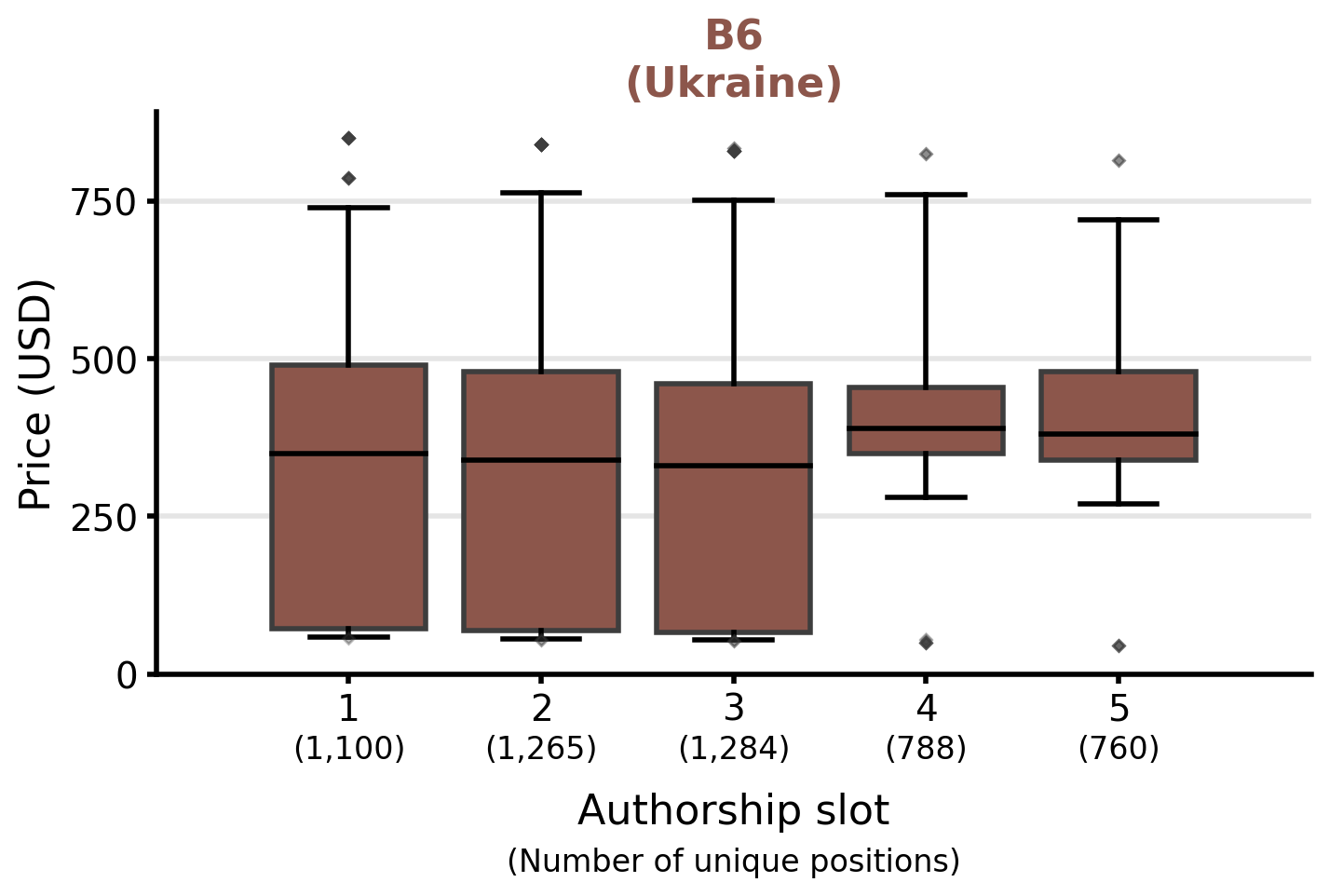}
    \caption{Distributions of prices for authorship positions on academic articles advertised by B6. Only the most recently-advertised price for a given authorship position is included. Boxplots show median as a horizontal line, interquartile range as boxes, 2.5th and 97.5th percentiles as whiskers, and outliers as diamonds.}
    \label{fig:b6_all}
\end{figure}

\begin{figure}[h!tbp]
    \centering
        \includegraphics[width=0.75\textwidth]{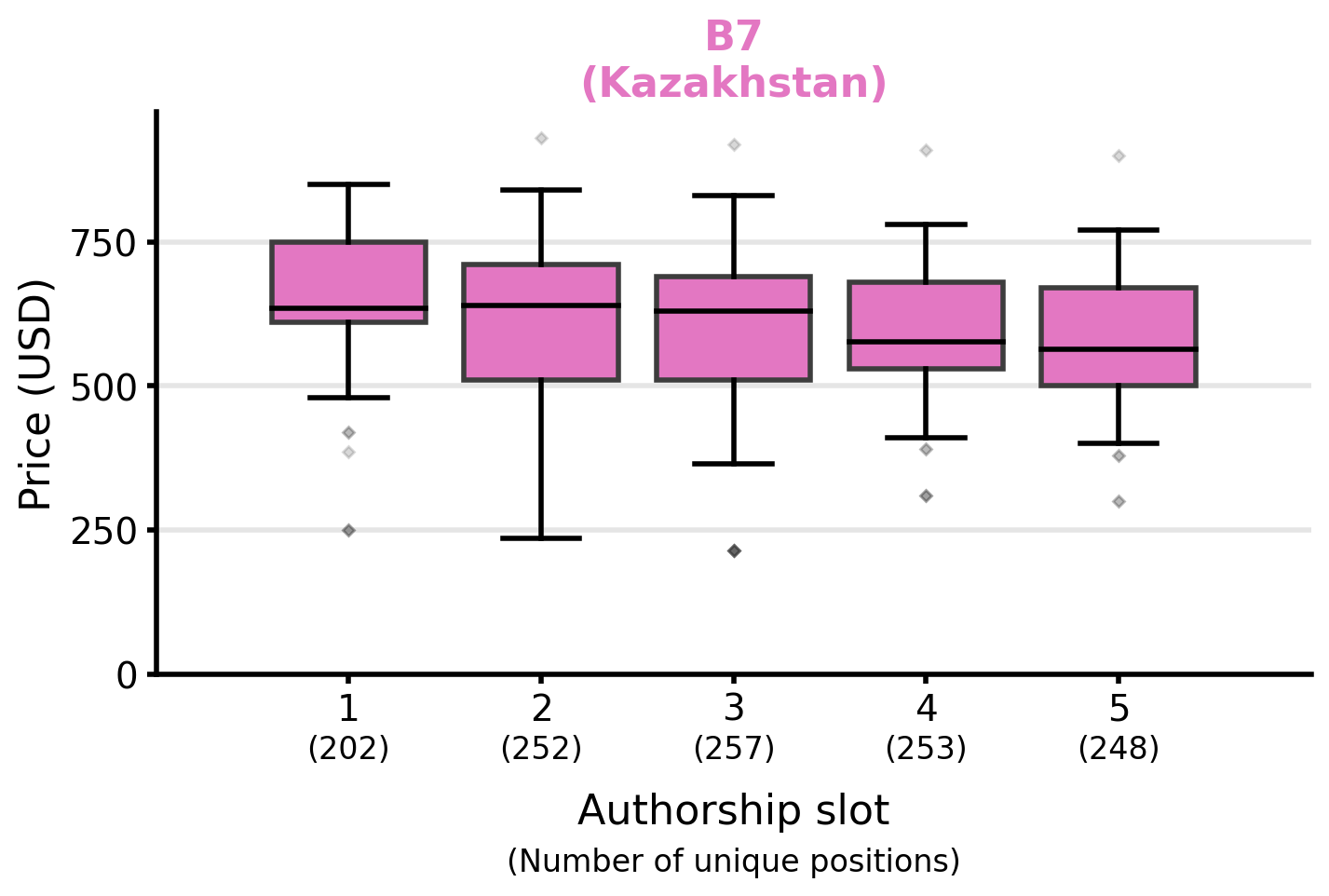}
    \caption{Distributions of prices for authorship positions on academic articles advertised by B7. Only the most recently-advertised price for a given authorship position is included. Boxplots show median as a horizontal line, interquartile range as boxes, 2.5th and 97.5th percentiles as whiskers, and outliers as diamonds.}
    \label{fig:b7_all}
\end{figure}

\end{document}